\documentclass[prl,superscriptaddress,showpacs,twocolumn]{revtex4-1}

\usepackage[english]{babel}
\usepackage[unicode]{hyperref}
\usepackage{amsmath}
\usepackage{amsfonts}
\usepackage{amssymb}
\usepackage{graphicx}
\usepackage{epstopdf}
\usepackage{color}
\usepackage{bm}

\newcommand{\be}{\begin{eqnarray}}
\newcommand{\ee}{\end{eqnarray}}
\newcommand{\ba}{\begin{array}}
\newcommand{\ea}{\end{array}}

\begin{document}

\title{Dephasing time in graphene due to interaction with flexural phonons}

\author{Konstantin S. Tikhonov *}
\email{tikhonov@physics.tamu.edu}
\affiliation{Department of Physics \& Astronomy, Texas A\&M University, College Station, TX 77843-4242, USA}
\affiliation{Department of Condensed Matter Physics, The Weizmann Institute of Science, 76100 Rehovot, Israel}
\affiliation{L. D. Landau Institute for Theoretical Physics, 117940 Moscow, Russia}
\affiliation{Moscow Institute of Physics and Technology, 141700 Moscow, Russia}

\author{Wei L.Z. Zhao *}
\email{wei.zhao@tamu.edu}
\affiliation{Department of Physics \& Astronomy, Texas A\&M University, College Station, TX 77843-4242, USA}
\affiliation{Department of Condensed Matter Physics, The Weizmann Institute of Science, 76100 Rehovot, Israel}

\author{Alexander M. Finkel'stein}
\affiliation{Department of Physics \& Astronomy, Texas A\&M University, College Station, TX 77843-4242, USA}
\affiliation{Department of Condensed Matter Physics, The Weizmann Institute of Science, 76100 Rehovot, Israel}
\affiliation{Institut f{\"u}r Nanotechnolsogie, Karlsruhe Institute of Technology, 76021 Karlsruhe, Germany}


\date{\today}

\begin{abstract}
We investigate decoherence of an electron in graphene caused by
electron-flexural phonon interaction. We find out that flexural phonons
can produce dephasing rate comparable to the electron-electron one. The
problem appears to be quite special because there is a large interval of
temperature where the dephasing induced by phonons can not be obtain using the golden rule.
We evaluate this rate for a wide range of density ($n$) and temperature ($T$) and determine several
asymptotic regions with temperature dependence crossing over from $\tau_{\phi
}^{-1}\sim T^{2}$ to $\tau_{\phi}^{-1}\sim T$ when temperature increases. We
also find $\tau_{\phi}^{-1}$ to be a non-monotonous function of $n$. These
distinctive features of the new contribution can provide an effective way to
identify flexural phonons in graphene through the electronic transport by
measuring the weak localization corrections in magnetoresistance.
\end{abstract}

\maketitle

\emph{Introduction.}
The transport properties of graphene have attracted much attention \cite{SarmaRMP}
since the first discovery of this fascinating material \cite{Neto09rmp}. It is
promising for various applications due to its high charge mobility and unique
heat conductivity. Theoretically, it was realized long ago
\cite{Mariani,Ochoa11,Gornyi12} that these transport properties of
free-standing (suspended) graphene are strongly influenced by flexural
(out-of-plane) vibrational modes that deform the graphene sheet. From the
experimental point of view, the effect of flexural phonons (FPs) was clearly observed in heat transport
\cite{Balandin08,Seol10}. However, it is a more challenging task to identify
the effect of flexural phonons in electronic transport
\cite{Bolotin08,Castro10}. This is because the contribution of electron-phonon
interactions to momentum relaxation remains small even at high
temperatures, with the main source of the relaxation being elastic impurities
\cite{Hwang}.

\begin{figure}[ptb]
\includegraphics[width=180pt]{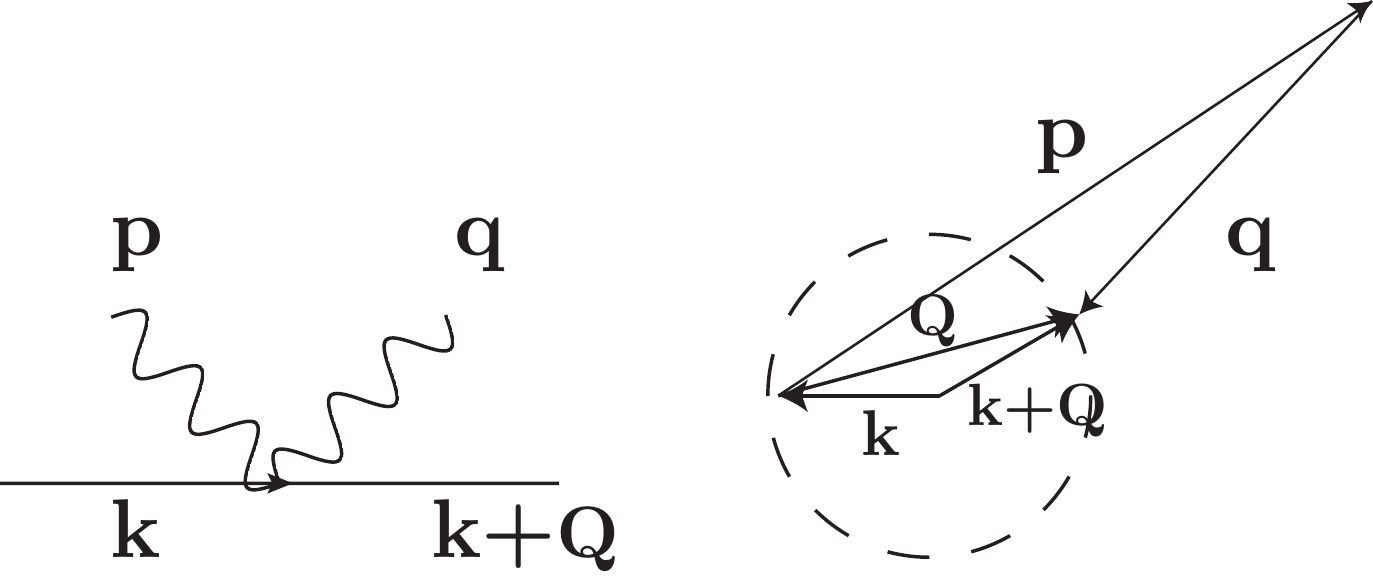}\caption{On the left:
scheme of the el-FP interaction process, where the solid line represents an
electron, and the wavy lines represent FPs. On the right: FPs can have momenta $\mathbf{p}$,
$\mathbf{q}$ much larger than the transferred momentum $\mathbf{Q}.$ Under the
conditions discussed in the paper, the scattering process is considered as
semi-elastic.}%
\label{FLpic}%
\end{figure}

The dephasing rate $\tau_{\phi}^{-1}$, on the other hand, is a more suitable quantity for studying
FPs, since static impurities do not cause dephasing. Usually, electron-electron
interactions, \cite{AAK82,Wu07,Tikhonenko09,Lundeberg10,Jobst12} are
considered the primary mechanism for dephasing. In this letter we discuss
dephasing caused by the electron-flexural phonon (el-FP) interaction in
graphene. It is the softness of the flexural mode and the coupling of an
electron to two FPs simultaneously (see Fig. \ref{FLpic} for illustration)
that make the contribution of FPs to $\tau_{\phi}^{-1}$ significant in a suspended sample,
and at large enough densities comparable with the one
caused by the electron-electron interaction. Because of the quadratic spectrum
of FPs, $\omega_{k}=\alpha k^{2}$, they are much more populated as compared
with in-plane phonons. In addition, the coupling to two FPs considerably
increases the phase space available for inelastic processes as compared to
the interaction with a single phonon. The point is that in graphene the Fermi
momentum, $k_{F},$ is relatively small. As a result, the interaction of a
single phonon with electrons is determined by the Bloch-Gr{\"{u}}neisen
temperature, $T_{BG}\sim\omega_{2k_{F}}$, rather than the temperature, when
$T\gg T_{BG}$ \cite{Efetov10}. In such a case, one needs to exploit other
scattering mechanisms to overcome the limitations induced by the smallness of
$k_{F}$ \cite{Song12}. In the case of el-FP interaction, coupling to two phonons radically
changes the situation. Now only the transferred momentum should be small,
while individually a\ FP may have a momentum much larger than $k_{F},$ up to
the thermal momentum $q_{T}.$

Still, as we shall demonstrate, the problem of dephasing due to the el-FP
interaction appears to be quite special, because the softness of FPs, i.e.
unique smallness of $T_{BG}$, leads to the existence of a temperature range
where dephasing rate cannot be obtained using the golden rule (GR). Rather,
both the self-energy and the vertex processes \cite{vonDelft07} should be
treated simultaneously. This results in a transition from $\tau_{\phi}%
^{-1}\sim T^{2}$ to $T$ with increasing temperature for the dephasing rate
induced by FPs.

\emph{The electron-flexural phonon interaction.}
Lattice dynamics of the single-layer graphene can be described in terms of
the displacement vector $\mathbf{u=}\left(  u_{x},u_{y},h\right)  $
\cite{Chaikin}. Here $u_{x,y}$ describe the in-plane modes, while the
out-of-plane displacement $h$ describes the flexural mode. The displacement
vector leads to a non-linear strain tensor
$u_{ij}=\frac{1}{2}\left(\partial_{i}u_{j}+\partial_{j}u_{i}+\partial_{i}h\partial_{j}h\right),$
where $\left(  i,j\right)  =\left(  x,y\right)  $ are spatial indices. The
lattice modes interact with electrons through emergent scalar and vector
potential fields \cite{Ando02, Manes07}:
\begin{align}
\varphi &  =g_{1}\left(  u_{xx}+u_{yy}\right)  ,\nonumber\\
~\mathbf{A}^{\alpha}  &  =s^{\alpha}g_{2}/v_{F}\left(  u_{xx}-u_{yy}%
,-2u_{xy}\right)  , \label{fields}%
\end{align}
where $g_{1}=30$eV, $g_{2}=7.5$eV \cite{Castro10} and $v_{F}$ is the
Fermi velocity. Index $\alpha=K,K^{\prime}$ describes two valleys of the conducting electron
band, and factor $s^{K/K^{\prime}}=\pm1$ reflects the fact that the emergent
vector potential $\mathbf{A}^{\alpha}$ respects the time reversal symmetry.

Thermal fluctuations of the lattice produce variations in the potentials.
Averaging over lattice vibrations one finds the correlation functions of the
potentials as
\begin{align}
\left\langle \varphi\left(  \mathbf{Q},\Omega\right)  \varphi\left(
-\mathbf{Q},-\Omega\right)  \right\rangle  &  =\mathcal{\phi}\left(
\mathbf{Q},\Omega\right)  ,\nonumber\\
\left\langle A_{i}^{\alpha}\left(  \mathbf{Q},\Omega\right)  A_{j}^{\beta
}\left(  -\mathbf{Q},-\Omega\right)  \right\rangle  &  =s^{\alpha}s^{\beta
}\mathcal{A}_{ij}\left(  \mathbf{Q},\Omega\right)  . \label{hh}%
\end{align}
To proceed, we introduce the correlation function for FP%
\begin{equation}
\left\langle h\left(  \mathbf{k},\omega\right)  h\left(  -\mathbf{k}%
,-\omega\right)  \right\rangle \equiv H\left(  k\right)  2\pi\delta\left(
\omega-\omega_{k}\right)  ,
\end{equation}
where $H\left(k\right)  =\frac{n\left(  \omega_{k}\right)  }{\rho\omega_{k}}$. In this equation,
$n\left(  \omega\right)  $ is the Planck distribution function and $\rho$ is
the mass density of the graphene sheet. One can propose the following form of the
spectrum of the flexural phonon:
\begin{equation}
\omega_{k}=\alpha k^{2}\Theta\left(  k\right)  ,\text{ }\Theta\left(
k\right)  =\sqrt{1+Z^{-1}\left(  q_{c}/k\right)  ^{\eta}}, \label{phonon}%
\end{equation}
where $\Theta\left(  k\right)  $ describes a transition from the bare spectrum
at high momentum to the renormalized spectrum $\sim k^{2-\eta/2}$ in the low
momentum limit. At $k<q_{c}\left(T\right)  =\frac{\sqrt{T\Delta_{c}}}{v_{F}}$
the quadratic spectrum for
the flexural mode ceases to work due to anharmonicity. Here $\Delta_{c}%
\approx18.7$eV \cite{Gornyi12} reflects the energy scale of anharmonicity.
The anharmonicity is related to the $h^{4}$-vertex,
arising as a result of integrating out fast $u$-modes, which are coupled to
$h$-mode \cite{Nelson}. Below we will exploit the value $Z\sim2$, and take $\eta\approx0.8$
from the numerical solution of the self-consistent screening approximation theory \cite{Radzihovsky92, Zakhachenko}.

We consider graphene away from the Dirac point at chemical potential $\mu\gg
T$. Besides $k_{F},$ the relevant momentum scales in the problem are: thermal
momentum $q_{T}=\sqrt{T/\alpha}\approx0.05\sqrt{T[\text{K}]}$/$%
\operatorname{\text{\AA}}%
$, and $q_{c}(T)\approx0.01\sqrt{T[\text{K}]}/%
\operatorname{\text{\AA}}%
$ which signals the transition to the renormalized FP spectrum. From now on,
we will concentrate on the realistic situation from the experimental
viewpoint: $k_{F}\ll q_{T}$, i.e., $T\gg T_{BG}$. The Bloch-Gr{\"{u}}neisen
temperature $T_{BG}=\omega_{2k_{F}}\approx0.4\Theta\left(  2k_{F}\right)  n$K,
where $n$ is the electronic density measured in units of $10^{12}$%
cm$^{-2}$. Note $T_{BG}$ is extraordinarily small for all relevant densities.
As we have already emphasized, see Fig. \ref{FLpic}, the momentum transfer in
the el-FP interaction is limited by $2k_{F}$. Nevertheless, the extended
structure of the correlation functions $\mathcal{\phi}\left(\mathbf{Q},\Omega\right)$ and
$\mathcal{A}_{ij}\left(\mathbf{Q},\Omega\right)$ enables electrons to have energy transfer exceeding
the phonon energy $\omega_{2k_{F}}$.

Main tool to probe electronic coherence is magnetoresistance
\cite{Altshuler80}, which gives a direct access to the weak localization
corrections to conductivity, controlled by the dephasing rate $\tau_{\phi
}^{-1}$. The weak localization correction to conductivity in graphene can be
written as \cite{Altshuler81,McCann06}%
\begin{equation}
\Delta\sigma=-\frac{2e^{2}D}{\pi}\sum_{l}\int dtC^{l}\left(  -t/2,t/2\right)
,
\end{equation}
where $l$ sums over four Cooperon channels relevant for the magnetoresistance.
Physically, $C^{l}\left(  -t/2,t/2\right)  $ represents the interference of a
pair of time reversed trajectories in the channel $l$ that start at $-t/2$ and
return to the initial point at $t/2$.
More generally, the Cooperon matrix $C_{s_{1}s_{2}}^{l_{1}l_{2}}$ is labelled
by two isospin numbers $s_{1,2}$ and two pseudospin numbers $l_{1,2}$. This matrix is
diagonal in the pseudospin space even in the presence of interactions that
preserve sublattice and valley indices. The Cooperon channels relevant for
magnetoresistance are the isospin singlets, $C^{l}\equiv C_{00}^{ll}$ $\left(
l=0,x,y,z\right)$, that do not have gaps comparable with $\tau^{-1}$, the
elastic scattering rate due to impurities. Therefore, we restrict ourselves to
this subspace.

\begin{figure}[h]
\centering\includegraphics[width=0.4\textwidth]{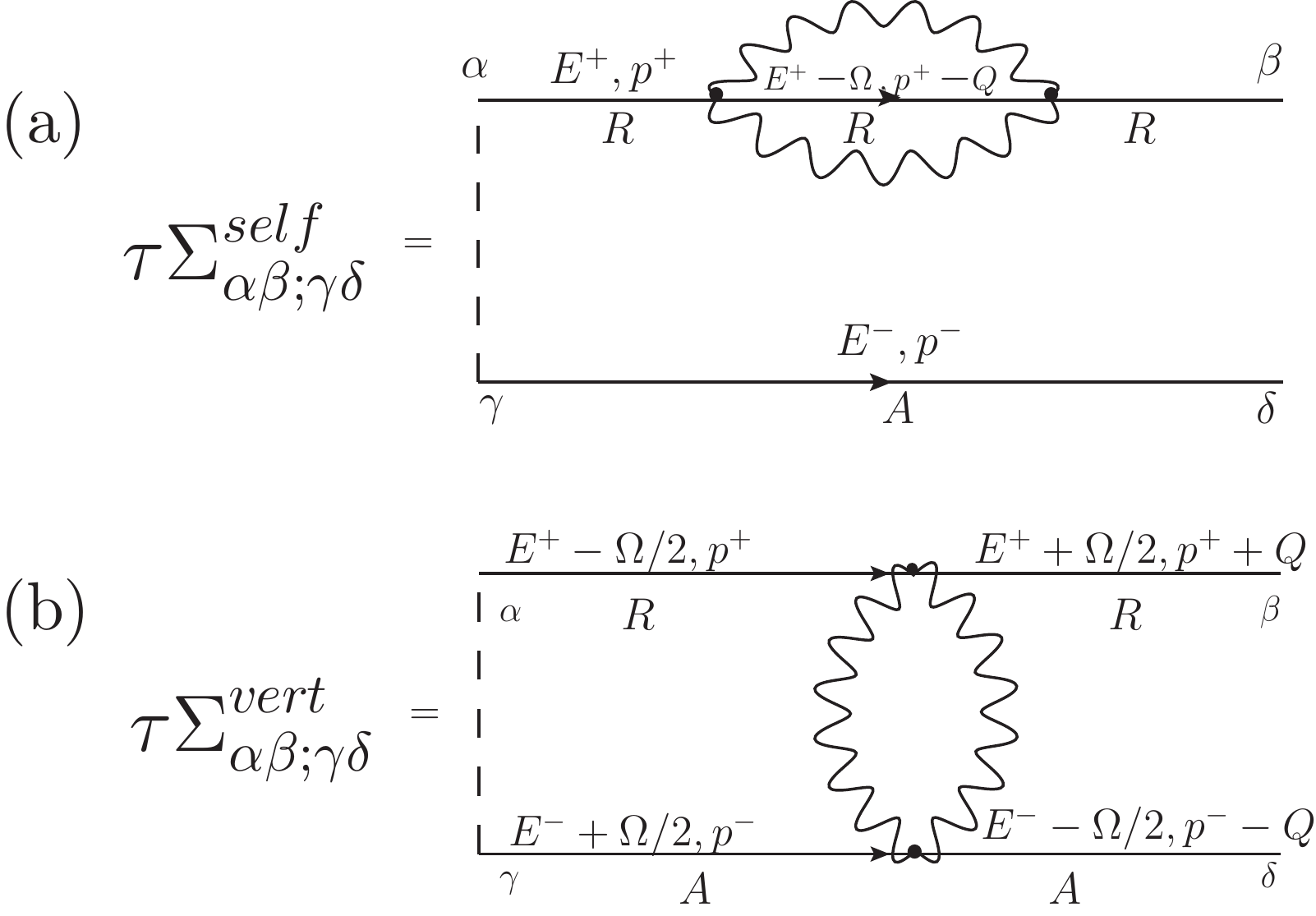}
\caption{Diagrammatic representation of (a) the self-energy and (b) the
vertex FP-contribution to the Cooperon; see also Fig. 2c in \cite{Supp}.}
\label{diag}
\end{figure}

To include el-FP interaction into the Cooperon, one can write down a
Bethe-Salpeter equation for a particular Cooperon channel $C^{l}$, see
Fig. \ref{diag}. In the
following we will not solve the equation exactly, but instead, we will
estimate the upper bound of the Cooperon decay rate \cite{Eiler84,Marquardt07}.
 We start by writing down an ansatz that reads as \cite{vonDelft07}
\begin{equation}
C^{l}\left(t_{1},t_{2}\right) = C_{0}^{l}\left(t_{1}-t_{2}\right)
e^{-F^{l}\left(  t_{1},t_{2}\right)  }.
\end{equation}
Here $C_{0}^{l}\left(  t\right)  $ is the diffusion propagator describing the
bare Cooperon, and $F^{l}\left(  t_{1},t_{2}\right)  $ is a decay function
characterizing the effect of the el-FP interaction \cite{Supp}.

\emph{Dephasing due to scalar potential fluctuations.} For the scalar potential
correlation function one obtains:

\begin{align}
\mathcal{\phi}\left(  \mathbf{Q},\Omega\right)  =  &  \frac{1}{8}g_{1}%
^{2}\left(  Q\right)  \int\left(  d^{2}\mathbf{p}\right)  \left(
d^{2}\mathbf{q}\right)  \left[  \mathbf{p}\cdot\mathbf{q}\right]
^{2}\nonumber\\
\times &  H\left(  p\right)  H\left(  q\right)  \delta_{\mathbf{p,q}}\left(
\Omega,\mathbf{Q}\right)  , \label{phicorr}%
\end{align}
where $\delta_{\mathbf{p,q}}\left(  \Omega,\mathbf{Q}\right)  \equiv\sum_{\pm
}\left(  2\pi\right)  ^{3}\delta\left(  \Omega\pm\omega_{\mathbf{p}}\pm
\omega_{\mathbf{q}}\right)  \times$ $\delta\left(  \mathbf{Q}-\mathbf{p-q}%
\right)  $, and $\omega_{\mathbf{p,q}}$ are given by Eq. (\ref{phonon}). Here
summation includes four different processes of emission/absorption of two
FPs by an electron. The screened coupling constant $g_{1}\left(  Q\right)
=g_{1}\frac{Q}{Q+\varkappa}$, where $\varkappa=g_{e}Nk_{F}$, $N=4$ is the
spin-valley degeneracy in graphene, and $g_{e}\sim1$ describes the
renormalized Coulomb interaction \cite{Kotov}. Since each time an electron is
coupled to \textit{two} flexural phonons, $\mathcal{\phi}$ describes a phonon
loop and, therefore, in the momentum-frequency domain $\mathcal{\phi}\left(
\mathbf{Q},\Omega\right)  $ has a extended support rather than a $\delta
$-function peak. As a result, the decay function for the scalar potential
$F_{\phi}\left(t\right)$
(which is the same for all channels) can be expressed as a convolution of the three
factors \cite{Supp}: i) the correlation
function $\mathcal{\phi}\left(  \mathbf{Q},\Omega\right)  $, ii) function
$\mathcal{B}_{\phi}\left(\mathbf{Q}\right)  $, describing the ballistic electron's motion,
and iii) factor $\mathcal{C}^{\phi}\left(  \Omega,t\right)  $, reflecting the
relation between the self-energy and vertex diagrams:%
\begin{equation}
F_{\phi}\left(  t\right)  =t\int\left(  d\mathbf{Q}\right)  \left(
d\Omega\right)  \mathcal{\phi}\left(  \mathbf{Q},\Omega\right)  \mathcal{B}_{\phi}
\left(  \mathbf{Q}\right)  \mathcal{C}^{\phi}\left(  \Omega,t\right)  .
\label{ft}%
\end{equation}
Here,
\begin{equation}
\mathcal{B}_{\phi}\left(  \mathbf{Q}\right)  =\frac{2}{v_{F}Q}\left(  1-\left(
Q/2k_{F}\right)  ^{2}\right)  ^{1/2}\theta\left(  2k_{F}-Q\right)  ,
\end{equation}
where the Heaviside theta function $\theta\left(  2k_{F}-Q\right)  $ restricts
momentum that can be exchanged between FPs and electrons.
The factor $\mathcal{C}^{\phi}\left(  \Omega,t\right)  $ is equal to
\begin{equation}
\mathcal{C}^{\phi}\left(  \Omega,t\right)  =1-\frac{\sin\Omega t}{\Omega t},
\label{sinfac}%
\end{equation}
and it describes the balance between the self-energy
and vertex diagrams on Fig. \ref{diag}. $\mathcal{C}^{\phi}$ is sensitive to dynamic aspect of the scattering
event and, because of this, alters temperature dependence of $\tau_{\phi
}^{-1}$.

\begin{figure}[ptb]
\includegraphics[width=200pt]{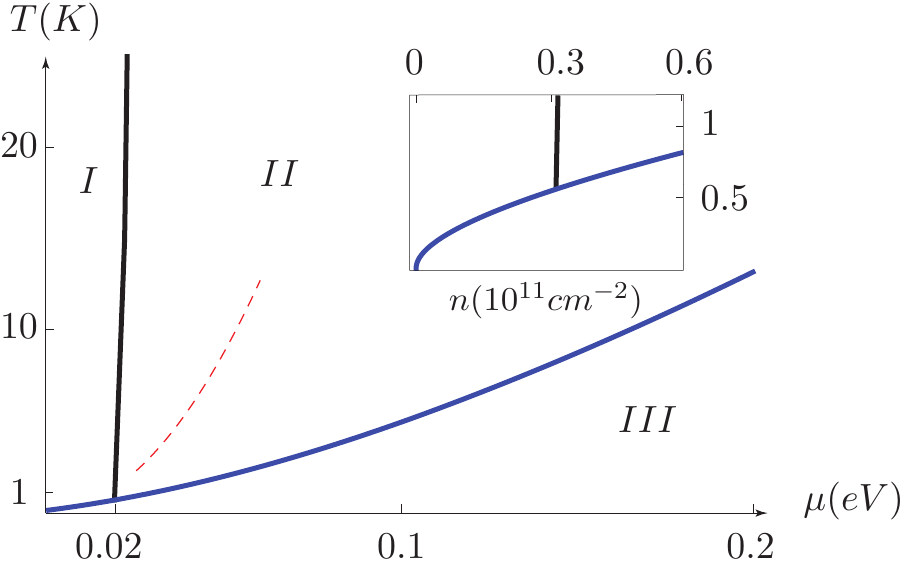}\caption{Phase diagram of
the dephasing rate due to FPs with scalar coupling. The blue and black lines
divide the whole $\left(  T-\mu\right)  $ plane into three regions, see the
text for explanations. The blue line coincides with the maximum of the
dephasing rate as a function of chemical potential at a fixed temperature, see
Fig. (\ref{density}). The red dashed line representing a fragment of $\xi=1$
is shown here for orientation. The inset is a zoom in of the intersection area
of the blue and black lines plotted as a function of the electronic density.}%
\label{phasediag}%
\end{figure}

The dephasing rate $\tau_{\phi}^{-1}$ is defined according to $F_{\phi}\left(
\tau_{\phi}\right)  =1$. The decay function can be most conveniently expressed
as
\begin{equation}
F_{\phi}\left(  t\right)  =c_{\phi}^{2}tTf\left(  \mathcal{T},\xi\right)
\frac{T}{\mu}, \label{scalarF}%
\end{equation}
where $c_{\phi}=\frac{g_{1}/\rho\alpha^{2}}{2\pi g_{e}N}\sim1.2$ is
dimensionless coupling constant and $f$ is a dimensionless function of two
parameters: $\mathcal{T}=\alpha k_{F}^{2}t$ and $\xi=Z^{-1/\eta}q_{c}/k_{F}$
\cite{Supp}. Parameter $\xi$ originates from the renormalization of the FP
spectrum described by $\Theta$ in Eq. (\ref{phonon}); $\Theta\left(
k_{F}\right)  =\sqrt{1+\xi^{\eta}}.$ At small $\mathcal{T}$ the function $f$
is linear in $\mathcal{T}$, and it saturates at $\mathcal{T}\gg1\mathcal{.}$

The results are illustrated with the help of Fig. \ref{phasediag}, where
regions I, II and III with a different dephasing rate behavior are indicated
in the $\left(  T-\mu\right)  $ plane. The regions are divided in accord with
the importance of the renormalized spectrum of the FP and the relative
contributions of the self-energy and vertex diagrams. In region I, which is on
the left of the black line (i.e., at small densities), the characteristic
momenta of $p$ and $q$ in Eq. (\ref{phicorr}) do not exceed $q_{c}.$
Therefore, the renormalization of the FP spectrum is important, and
$\omega_{q}\sim$ $q^{2-\eta/2\text{ }}$ should be used \cite{Vertex}. In region II, since
the characteristic momenta of the FPs are larger than $q_{c}$, it suffices to
use the quadratic spectrum for FPs. In region III, which is in the bottom part
below the blue line, the dephasing time is long and only the self-energy
diagram is important. Hence, the factor $\mathcal{C}^{\phi}$ reduces to $1$,
and dephasing rate coincides with the out-scattering rate, $\tau_{out}^{-1}$,
obtained from the golden rule \cite{Gornyi12}. (In this calculation, $q_{c}$
just provides an infrared cut-off.) Above the blue line, in regions I and II,
both the self-energy and vertical diagrams are relevant, and the factor
$\mathcal{C}^{\phi}\left(  t\right)  $ is important; see also \cite{Montambaux05}.
Due to the two-phonon structure of the correlation function
of the FP pairs participating in the inelastic process, the influence of this
factor on the dephasing rate is rather non-trivial, so that one cannot expand
$\mathcal{C}^{\phi}\left(  t\right)  $.

In Fig. \ref{phasediag}, the blue and black lines have been found by matching
the asymptotic behavior \cite{Supp} of the dephasing rates deep in regions I,
II and III. We introduce $\left(  \mu_{0},T_{0}\right)  ,$ the values of the crossing
point of the blue and black lines as characteristic scales: $\mu_{0}\sim
\frac{\gamma}{c_{\phi}^{2}}\Delta_{c}$ and $T_{0}\sim\frac{\gamma}{c_{\phi
}^{2}}\mu_{0}$. Here, we have introduced $\gamma=\frac{\alpha\Delta_{c}}%
{v_{F}^{2}}\sim0.02$, which is a parameter describing the adiabaticity of the
el-FP interaction. Under a given choice of parameters, it can be found
numerically that $\mu_{0}\approx0.02$eV and $T_{0}\approx0.6$K. The dephasing
rate in different regions can be expressed as
\begin{equation}
\tau_{\phi}^{-1}\left(  T\right)  =\gamma T\times\left\{
\begin{array}
[c]{cc}%
0.48\left(  \mu/\mu_{0}\right)  ^{\frac{4-\eta}{8-5\eta}} & \text{I}\\
0.18\sqrt{\mu/\mu_{0}} & \text{II}\\
0.24\frac{T/T_{0}}{\mu/\mu_{0}}\log\xi^{-1} & \text{III.}%
\end{array}
\right.  \label{main}%
\end{equation}
These expressions are obtained using asymptotic behavior of the function $f$
in Eq. (\ref{scalarF}) and, therefore, are only applicable far away from the
borderlines. At low enough temperatures, $\mathcal{C}^{\phi
}\left(  t\right)  =1$ and the function $f\left(  \mathcal{T},\xi\right)  $ is
independent of $\mathcal{T}$. Hence, $\tau_{\phi}^{-1}\sim T^{2}$ in region
III, which is a GR result. At high temperatures the phonons contributing to
the electronic dephasing become quasi-static and, consequently, the dephasing
rate is smaller than the out-scattering rate $\tau_{out}^{-1}$. Unlike region III, in regions I and
II the dephasing rate is determined by a non-GR expression, and is
proportional to temperature, irrespective of $\eta$. The existence of the linear in $T$ regime is the main result of our paper.

\begin{figure}[tbh]
\centering
\includegraphics[width=200pt]{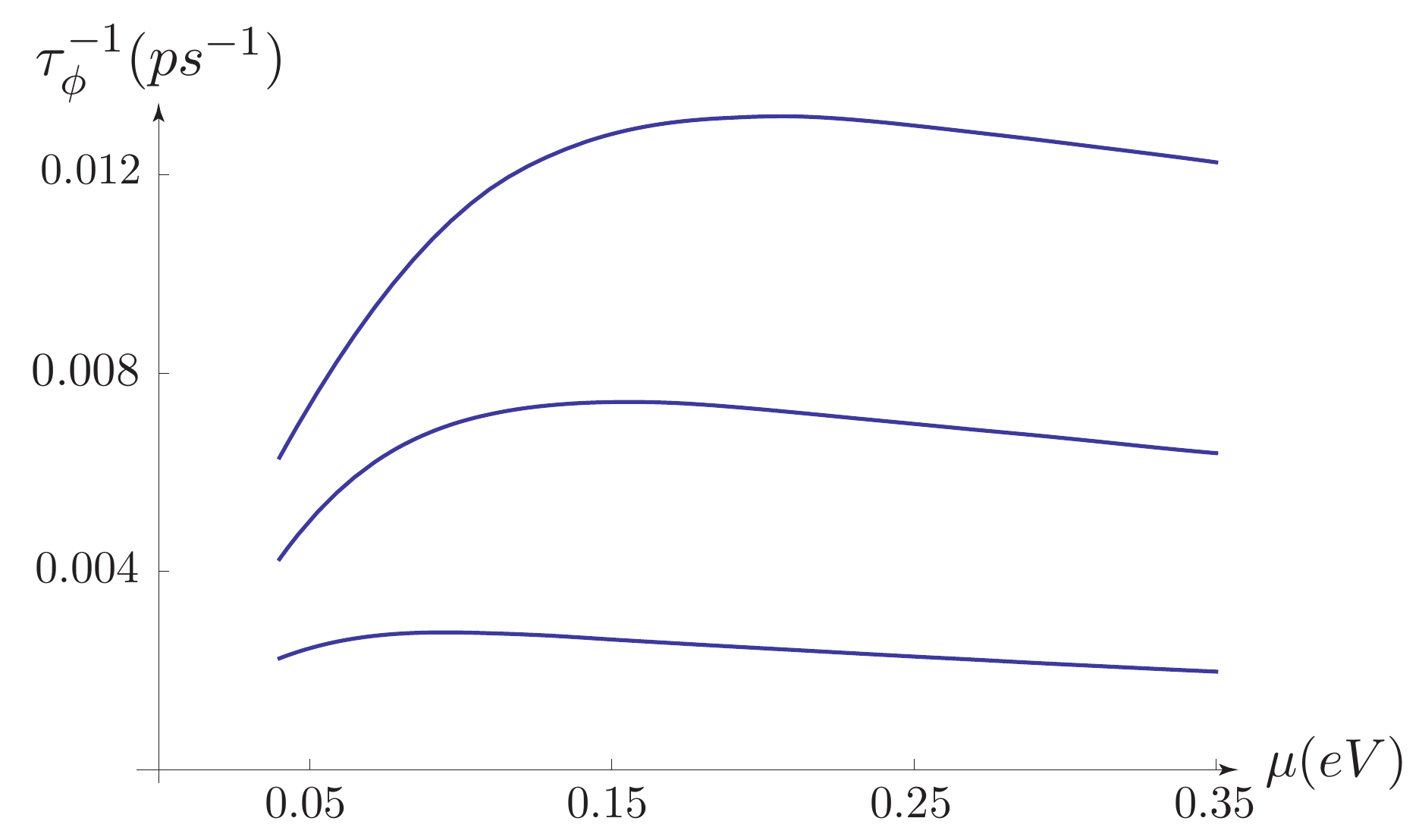}\caption{Dephasing rate
as a function of the chemical potential at different temperatures. From top to
buttom: T=15K, 10K, 5K.}%
\label{density}%
\end{figure}

By comparing the rates in regions II and III, one may conclude that there
should be a maximum in the dephasing rate as a function of $\mu$.\ Indeed, as
it is illustrated by Fig. \ref{density} such a maximum exists. The line
indicating the maximum essentially overlaps with the borderline between the
regions I, II and the region III, which is illustrated by the blue line in
Fig. \ref{phasediag}.

\emph{Dephasing due to vector potential fluctuations.} Unlike the scalar potential,
the dephasing rates induced by vector potential are different for different channels owing to the factor
$\mathcal{C}_{l}^{A}\left(  \Omega,t\right)  =1+s_{l}\frac{\sin\omega t}{\omega t},$ where $s_{l}=-1$
 for the intervalley Cooperons
$\left(l=0,z\right)$ and $s_{l}=1$ for the intravalley Cooperons $\left(l=x,y\right)$.
For the intervalley channels, the only relevant
for the magnetoresistance at weak fields, the dephasing rate produced by the vector
potential coupling is quite similar to its scalar counterpart,
Eq. (\ref{main}), with obvious modifications due to the change in the coupling constant
 and absence of screening for the vector potential \cite{Supp}.

\emph{Discussion.} We have analysed the dephasing rate induced by
FPs in graphene, and evaluated it for a wide range of $n$
and $T$ (see Fig. \ref{phasediag}.) We determined several asymptotic regions
with temperature dependence evolving from $\tau_{\phi}^{-1}\sim T^{2}$ to
$\tau_{\phi}^{-1}\sim T$ when temperature increases. (See Fig. 4 in \cite{Supp} for an illustration of the temperature dependence of the dephasing rate.) The transition to linear
behavior in $T$ is related to the fact that at high temperatures phonons
become slow on the time-scale of $\tau_{\phi}$.

The measured dephasing rate in graphene is usually compared to the
contribution induced by the electron-electron interaction, $\tau_{ee}^{-1}$,
which is linear in $T$ for $T<1/\tau_{tr}$ \cite{AAK82}. However, the observed
rate \cite{Wu07,Tikhonenko09,Lundeberg10}, when it is linear in $T$, always
exceeds the theoretical estimation. In view of the linear dependence on $T$ of the FP's contribution to
dephasing, it is reasonable to compare its value with $\tau_{ee}^{-1}$. In
principle, it is a competition between two mechanisms, each determined by a small parameter: the adiabatic
parameter $\gamma$ and sheet resistance $\rho_{\square}$ measured in units
of the quantum resistance. We compare the dephasing rates at density $n=10^{12}cm^{-2}$ when the sheet
resistance $\approx0.5k\Omega$. Under these conditions, both parameters
$\gamma$ and $\rho_{\square}$ are of the same value.
 Combining the
contributions arising from the scalar and vector potentials, we obtain
$\tau_{FP}^{-1}/\tau_{ee}^{-1}\approx0.2$.

The in-plane phonons generate
a dephasing $\tau_{in}^{-1}$ that at $T<T_{BG}^{in}$ is negligible compared
with $\tau_{FP}^{-1}$, while at $T>T_{BG}^{in}$ the rate $\tau_{in}^{-1}\sim T$
is comparable with $\tau_{FP}^{-1}$. (Note that for in-plane phonons, a region of non-GR dephasing rate, analogous to region II,
develops at temperatures $\gtrsim\mu$ that is too high to be relevant.) It is important that each of the three rates
$\tau_{ee}^{-1}$, $\tau_{FP}^{-1}$, and $\tau_{in}^{-1}$, has a distinct dependence
on the chemical potential.
While $\tau_{ee}^{-1}$ decreases with density, $\tau_{FP}^{-1}\varpropto
\mu^{1/2}$ and $\tau_{in}^{-1}\varpropto\mu$. This opens a way to identify
each of these mechanisms by studying the magnetoresistance as a function of
the chemical potential.

In our consideration, we had in mind suspended graphene. However, our
result may also be relevant for supported samples so long as they are coupled
to the substrate by weak Van der Waals forces \cite{Geim13}. One
may expect that such a weak coupling does not provide an essential change in
the phonon spectrum. Indeed, it is known that the phonon spectrum in graphene
\cite{Nika09} and graphite \cite{Al-Jishi82} are practically identical for the
corresponding branches. FPs in supported
samples have been discussed recently in connection with the heat transport
measurements in Refs. \cite{Balandin08,Seol10}. Until now flexural phonons have been a delicate object to detect in electronic
transport. We propose here to observe them through weak-localization measurements.

\emph{Acknowledgements.} The authors gratefully acknowledge A. Dmitriev, I. Gornyi, V. Kachorovskii, D.
Khmelnitskii and A. Mirlin for the useful discussions and valuable criticism.
The authors thank the members of the Institut f{\"{u}}r Theorie der
Kondensierten Materie at KIT for their kind hospitality. A.F. is supported by
the Alexander von Humboldt Foundation. The work is supported by the Paul and
Tina Gardner fund for Weizmann-TAMU collaboration, and National Science
Foundation grant NSF-DMR-100675.

* K.Tikhonov and W.Zhao contributed equally to this work.

\bibliography{Flexural}

\begin{thebibliography}{38}
\bibitem [{Ver()}]{Vertex}%
  \BibitemOpen
  \href@noop {} {}\bibinfo {note} {We, however, neglect the $h^{4}$-vertex
  corrections.}\BibitemShut {Stop}%
\end{thebibliography}

\end{document}


\title {Supplemental Material for \\ "Dephasing time in graphene due to interaction with flexural phonons"}

\author{Wei L.Z. Zhao}
\affiliation{Department of Physics \& Astronomy, Texas A\&M University, College Station,
TX 77843-4242, USA}
\affiliation{Department of Condensed Matter Physics, The Weizmann Institute of Science,
76100 Rehovot, Israel}
\author{Konstantin S. Tikhonov}
\affiliation{Department of Physics \& Astronomy, Texas A\&M University, College Station,
TX 77843-4242, USA}
\affiliation{Department of Condensed Matter Physics, The Weizmann Institute of Science,
76100 Rehovot, Israel}
\affiliation{L. D. Landau Institute for Theoretical Physics, 117940 Moscow, Russia}
\affiliation{Moscow Institute of Physics and Technology, 141700 Moscow, Russia}
\author{Alexander M. Finkel'stein}
\affiliation{Department of Physics \& Astronomy, Texas A\&M University, College Station,
TX 77843-4242, USA}
\affiliation{Department of Condensed Matter Physics, The Weizmann Institute of Science,
76100 Rehovot, Israel}
\affiliation{Institut f{\"u}r Nanotechnolsogie, Karlsruhe Institute of Technology, 76021
Karlsruhe, Germany}
\date{\today}

\maketitle

\section{Diagrammatic calculation of the decay function}

The decay function $F^{l}$ defined in Eq. (6m) ('m' refers to the main text)
can be illustrated with the following diagram:
\begin{figure}[tbh]
\centering
\includegraphics[width=130pt]{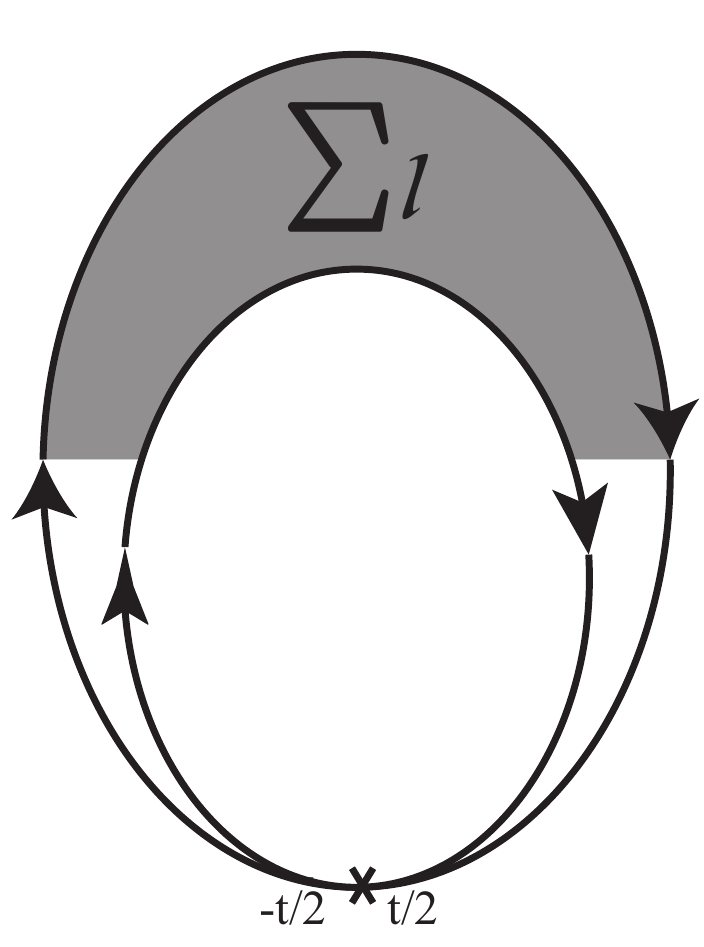}
\caption{Diagrammatic representation of the decay function $F\left( t\right) 
$. Here $\Sigma _{l}=\Sigma_{l}^{self}+\Sigma_{l}^{vert}$. }
\label{Cooperondiag}
\end{figure}

\begin{figure}[h]
\centering
\includegraphics[width=0.4\textwidth]{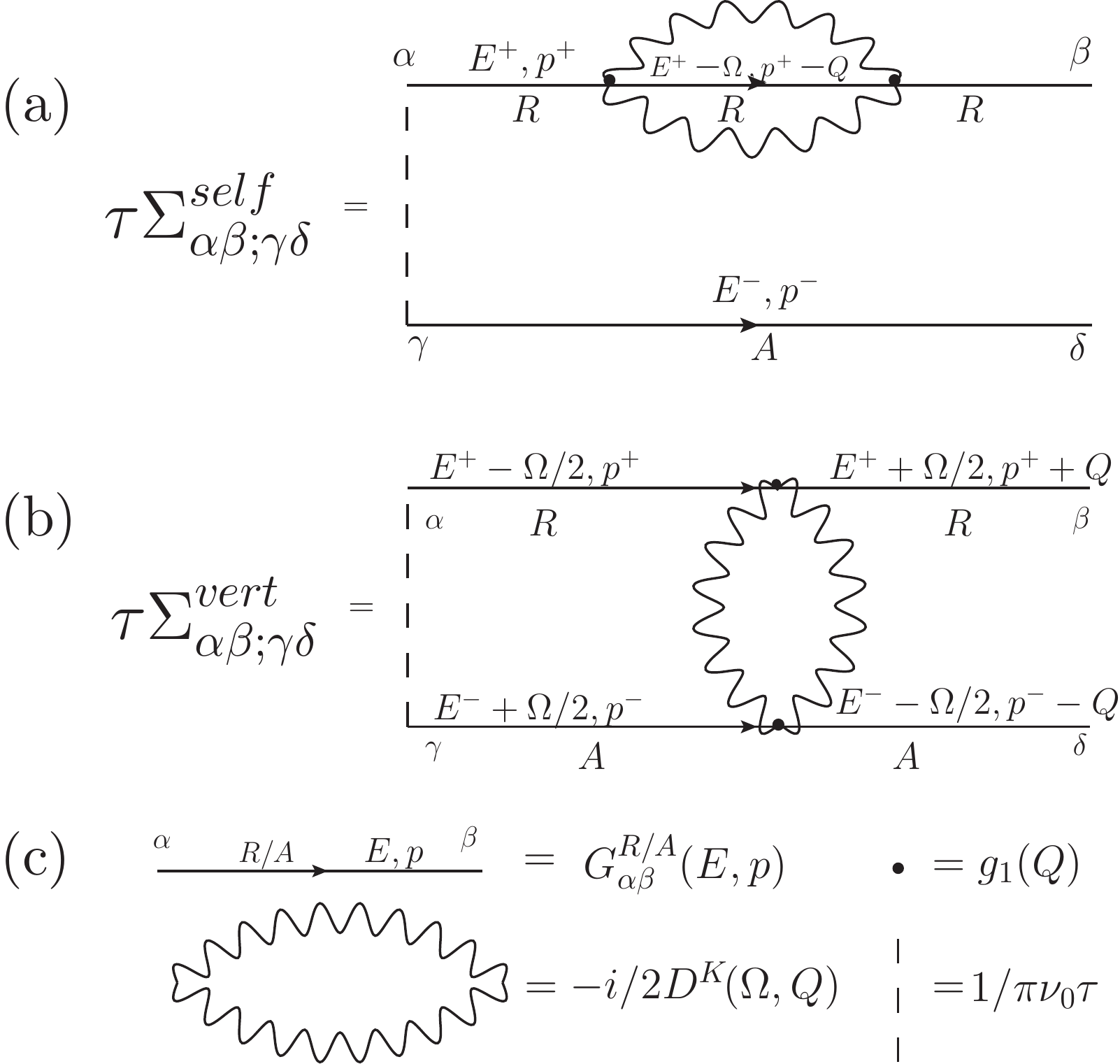}
\caption{Diagrammatic representation of (a) the self-energy contribution and
(b) the vertex contribution. (c) Diagrammatic dictionary for the various
objects involved: $E^{\pm}=\protect\mu\pm\tilde{\protect\omega}/2$ and $%
p^{\pm}=k_{F}\pm \tilde{q}/2$; $\protect\nu_{0}$ is the density of states at
Fermi energy, and $g_{1}\left(Q\right) $ is the coupling constant with
screening included.}
\label{diag}
\end{figure}

To the lowest order in the interaction propagators, the expression for $F^{l}
$ can be obtained as \cite{vonDelft07}%
\begin{equation}
F^{l}\left( t_{1},t_{2}\right) \simeq-\frac{C_{1}^{l}\left(
t_{1},t_{2}\right) }{C_{0}^{l}\left( t_{1}-t_{2}\right) },   \label{decay}
\end{equation}
where $C_{1}^{l}$ is the first order correction to the Cooperon in terms of
the interactions propagators. We perform the diagrammatic calculation with
the use of the matrix Greens function $G_{\alpha\beta}^{R/A}\left( \epsilon ,%
\mathbf{p}\right) =\frac{1}{2}\frac{\delta_{\alpha\beta}+\left( \mathbf{%
\Sigma}\cdot\mathbf{\hat{p}}\right) _{\alpha\beta}}{\epsilon\pm i/\left(
2\tau\right) -v_{F}p}$. Here we have introduced iso-pseudospin basis \cite%
{McCann06}, spanned by two sets of mutually commuting matrices, isospin-$%
\Sigma$ and pseudospin-$\Lambda$:%
\begin{align}
\Sigma_{0},\Sigma_{x} &
=\Pi_{z}\otimes\sigma_{x},\Sigma_{y}=\Pi_{z}\otimes\sigma_{y},\Sigma_{z}=%
\Pi_{0}\otimes\sigma_{z},  \notag \\
\Lambda_{0},\Lambda_{x} & =\Pi_{x}\otimes\sigma_{z},\Lambda_{y}=\Pi
_{y}\otimes\sigma_{z},\Lambda_{z}=\Pi_{z}\otimes\sigma_{0},
\end{align}
where $\Sigma_{0},\Lambda_{0}$ are unit matrices and $\Pi_{x,y,z}$ and $%
\sigma_{x,y,z}$ are Pauli matrices acting on valley and sublattice spaces
respectively. For each channel, the contribution of the el-FP interaction
can be separated into the scalar and vector potential ones. Thus, one can
write $F^{l}=F_{\phi}^{l}+F_{A}^{l}$. We concentrate first on dephasing
caused by the scalar potential fluctuation $\mathcal{\phi}\left( \mathbf{Q}%
,\Omega\right) $. The calculation for the vector potential contribution goes
along the similar lines, and results are presented in the end of this
Section. Due to the interaction propagators, one can write \cite{vonDelft07}%
\begin{align}
C_{1}^{l}\left( t_{1},t_{2}\right) = & \int\left( dQ\right) \left(
d\Omega\right) \left( d\tilde{q}\right) \left( d\tilde{\omega}\right) e^{-i%
\tilde{\omega}t_{12}}[C_{0}^{l}\left( \tilde{q},\tilde{\omega}\right)
\Sigma_{Q,\tilde{q}}^{l,self}\left( \tilde{\omega},\Omega\right)
C_{0}^{l}\left( \tilde{q},\tilde{\omega}\right)  \notag \\
& +e^{i\Omega\tau_{12}}C_{0}^{l}\left( \tilde{q},\tilde{\omega}%
-\Omega\right) \Sigma_{Q,\tilde{q}}^{l,vert}\left( \tilde{\omega}%
,\Omega\right) C_{0}^{l}\left( \tilde{q},\tilde{\omega}+\Omega\right) ], 
\label{C1}
\end{align}
where $t_{12}=t_{1}-t_{2}$ and $\tau_{12}=t_{1}+t_{2}$. Here, the effect of
the interaction is separated into the self-energy and vertex (vertical)
contributions, denoted as $\Sigma^{self/vert}$; see Fig. \ref{diag}. The
self-energy and vertex contribution of the channel $l$ are defined in the
iso-pseudospin basis via $\Sigma^{l,self/vert}\equiv\left(
\Sigma_{y}\Lambda_{y}\Lambda_{l}\right)
_{\alpha\beta}\Sigma_{\alpha\beta;\gamma\delta }^{self/vert,}\left(
\Sigma_{y}\Lambda_{l}\Lambda_{y}\right) _{\delta\gamma }$ (summation for
Greek letters is implied; indices $\phi\left( A\right) $ are omitted). Here,
the self-energy contribution, see Fig. \ref{diag}(a), can be written as

\begin{align}
\tau\Sigma_{\alpha\beta;\gamma\delta,}^{self,\phi}\left( \tilde{q},Q,\tilde{%
\omega},\Omega\right) & =-\frac{g\left( Q\right) ^{2}}{\pi \nu_{0}\tau}%
\int\left( d^{2}p\right) \left( -\frac{i}{2}D^{K}\left( Q,\Omega\right)
\right) \times  \notag \\
\{ & \left[ G^{R}\left( p^{+},E^{+}\right) G^{R}\left(
p^{+}-Q,E^{+}-\Omega\right) G^{R}\left( p^{+},E^{+}\right) \right]
_{\alpha\beta}G^{A}\left( p^{-},E^{-}\right) _{\gamma\delta}  \notag \\
+ & G^{R}\left( p^{+},E^{+}\right) _{\alpha\beta}\left[ G^{A}\left(
p^{-},E^{-}\right) G^{A}\left( p^{-}-Q,E^{-}-\Omega\right) G^{A}\left(
p^{-},E^{-}\right) \right] _{\gamma\delta}\},
\end{align}
where $E^{\pm}=\mu\pm\tilde{\omega}/2$, $p^{\pm}=k_{F}\pm\tilde{q}/2$. Note
that arguments $\tilde{q}$ and $\tilde{\omega}$ are related to the
propagation of the Cooperon, while $Q$ and $\Omega$ describe the
interactions with FPs causing the dephasing. The vertex contribution, see
Fig. \ref{diag}(b), is 
\begin{align}
\tau\Sigma_{\alpha\beta;\gamma\delta}^{vert,\phi}\left( \tilde{q},Q,\tilde{%
\omega},\Omega\right) & =-\frac{g\left( Q\right) ^{2}}{\pi \nu_{0}\tau}%
\int\left( d^{2}p\right) \left( -\frac{i}{2}D^{K}\left( Q,\Omega\right)
\right) \times  \notag \\
& \left[ G^{R}\left( p^{+},\varepsilon^{+}-\Omega\right) G^{R}\left(
p^{+}+Q,\varepsilon^{+}\right) \right] _{\alpha\beta}\left[ G^{A}\left(
p^{-},\varepsilon^{-}+\Omega\right) G^{A}\left( p^{-}-Q,\varepsilon
^{-}\right) \right] _{\gamma\delta},
\end{align}
where $\varepsilon^{\pm}=E^{\pm}\pm\Omega/2.$ Due to the softness of FPs,
the integrals in Eq. (\ref{C1}) converge at small frequencies $\Omega\ll T$.
Note that the assumption that $\Omega\ll T$ depends crucially on the fact
that $\tau_{\phi}^{-1}\ll T$ which has been checked a posteriori. Therefore,
when calculating the effect of interaction on the Cooperon propagators, one
may take only the classical (Keldysh) component of the interaction, $%
D_{\phi}^{K}\left( \mathbf{Q},\Omega\right) =-2i\mathcal{\phi}\left( \mathbf{%
Q},\Omega\right) $, and send $n\left( \omega_{k}\right) $ in $H\left(
k\right) $ to its classical limit, $\frac{T}{\omega_{k}}.$ If the full
quantum problem is considered, one needs to use $\frac{1}{\sinh\left(
\omega_{k}/T\right) },$ in order to incorporate Pauli principle due the
presence of other electrons \cite{Marquardt07}. This leads to the same
result for $T\gg\omega_{k},$ but ensures that quantum fluctuations do not
lead to dephasing at zero temperature.

In Eq. (\ref{C1}), the Cooperon variables $\left( \tilde{q},\tilde{\omega }%
\right) $ are small comparing to the electronic scales determining the el-FP
interaction process, i.e., $\tilde{q}\ll k_{F}$ and $\tilde{\omega}\ll\mu$.
Therefore, one can drop out $\left( \tilde{q},\tilde{\omega}\right) \ $%
dependences from the interaction propagators $\Sigma$. Furthermore, as far
as $T\gg T_{BG}$, the typical momentum transfer $Q\sim2k_{F}\gg1/v_{F}\tau$.
Therefore, dephasing can also be calculated assuming that electron's motion
during the interaction event is ballistic. After a simple calculation, one
concludes for scalar potential%
\begin{equation}
\Sigma_{\phi}^{l,self}=-\Sigma_{\phi}^{l,vert}=\Sigma_{\phi}\left(
Q,\Omega\right) ,
\end{equation}
where $\Sigma_{\phi}\left( Q,\Omega\right) \equiv\mathcal{\phi}\left( 
\mathbf{Q},\Omega\right) \mathcal{B}_{\phi}\left( Q\right) $. Here $\mathcal{%
\phi}\left( \mathbf{Q},\Omega\right) $ is defined in Eq. (7m), and $\mathcal{%
B}_{\phi}\left( Q\right) $ is defined in Eq. (9m). Plugging these results
back to Eq. (\ref{decay}) leads to Eq. (8m).

In the case of vector potential coupling, one calculates a\ diagram similar
to that in Fig. \ref{diag}, and gets%
\begin{equation}
\Sigma_{A}^{l,self}=s_{l}\Sigma_{A}^{l,vert}=\Sigma_{A}^{ij}\left(
Q,\Omega\right) \delta_{T}^{ij}\left( Q\right) ,
\end{equation}
where $\delta_{T}^{ij}\left( Q\right) =\delta^{ij}-Q^{i}Q^{j}/Q^{2}$ and $%
\Sigma_{A}^{ij}\left( Q,\Omega\right) =v_{F}^{2}\mathcal{A}^{ij}\left( 
\mathbf{Q},\Omega\right) \mathcal{B}_{A}\left( Q\right) .$ Here, the vector
potential correlation function%
\begin{align}
\mathcal{A}_{ij}\left( \mathbf{Q},\Omega\right) = & \frac{\mathbf{\hat{n}}%
_{i}\mathbf{\hat{n}}_{j}}{8}\left( g_{2}/v_{F}\right) ^{2}\int\left( d^{2}%
\mathbf{p}\right) \left( d^{2}\mathbf{q}\right) p^{2}q^{2}  \notag \\
\times & H\left( p\right) H\left( q\right) \delta_{\mathbf{p,q}}\left(
\Omega,\mathbf{Q}\right) ,   \label{aacorr}
\end{align}
where $\delta_{\mathbf{p,q}}\left( \Omega,\mathbf{Q}\right) =\sum_{\pm
}\left( 2\pi\right) ^{3}\delta\left( \Omega\pm\omega_{\mathbf{p}}\pm \omega_{%
\mathbf{q}}\right) \delta\left( \mathbf{Q}-\mathbf{p-q}\right) $ and the
summation includes four different processes of emission/absorption of two
FPs by an electron. Also, the factor describing the ballistic motion of the
electrons%
\begin{equation}
\mathcal{B}_{A}\left( Q\right) =\frac{2}{v_{F}Q}\left( 1-\left(
Q/2k_{F}\right) ^{2}\right) ^{-1/2}\theta\left( 2k_{F}-Q\right) .
\end{equation}
The corresponding decay function can be rendered as (compare with Eq. (8m))%
\begin{equation}
F_{A}^{l}\left( t\right) =t\int\left( d\mathbf{Q}\right) \left(
d\Omega\right) v_{F}^{2}\mathcal{A}^{ij}\left( \mathbf{Q},\Omega\right)
\delta_{T}^{ij}\left( Q\right) \mathcal{B}_{A}\left( Q\right) \mathcal{C}%
_{l}^{A}\left( \Omega,t\right) ,
\end{equation}
where%
\begin{equation}
\mathcal{C}_{l}^{A}\left( \Omega,t\right) =1+s_{l}\frac{\sin\Omega t}{\Omega
t},
\end{equation}
The intervalley Cooperons are coupled to the vector potential field of the
opposite signs. Thus, $s_{l}=\pm1$ for intra-/intervalley Cooperons. Further
on, one can resolve the transverse delta function $\delta_{T}^{ij}$ and get 
\begin{equation}
F_{A}^{l}\left( t\right) =t\int\sin^{3}\hat{Q}\left( d\mathbf{Q}\right)
\left( d\Omega\right) \mathcal{A}\left( \mathbf{Q},\Omega\right) \mathcal{B}%
_{A}\left( Q\right) \mathcal{C}_{l}^{A}\left( \Omega,t\right) ,
\end{equation}
where%
\begin{equation}
\mathcal{A}\left( \mathbf{Q},\Omega\right) =\frac{g_{2}^{2}}{8}\int\left(
d^{2}\mathbf{p}\right) \left( d^{2}\mathbf{q}\right) p^{2}q^{2}H\left( 
\mathbf{p}\right) H\left( \mathbf{q}\right) \delta_{\mathbf{p,q}}\left(
\Omega,\mathbf{Q}\right) .
\end{equation}

\section{Evaluation of the decay function}

\subsection{Scalar potential coupling}

Let us now evaluate the integral in Eq. (8m) explicitly. It is convenient to
use the time representation for the energy delta-function: 
\begin{equation}
\sum_{\pm}\left( 2\pi\right) \delta\left( \Omega\pm\omega_{\mathbf{p}%
}\pm\omega_{\mathbf{q}}\right) =4\int d\tau\cos\left( \omega_{p}\tau\right)
\cos\left( \omega_{q}\tau\right) \exp\left[ -i\Omega\tau\right] .
\end{equation}
After this, one can integrate Eq. (8m) in frequency $\Omega,\ $using%
\begin{equation}
\int\left( d\Omega\right) e^{-i\Omega\tau}\left( 1-\frac{\sin\Omega t}{%
\Omega t}\right) =\frac{1}{t}\Xi_{-}\left( \tau/t\right) ,
\end{equation}
where 
\begin{equation}
\Xi_{\pm}\left( s\right) \equiv\delta\left( s\right) \pm\frac{1}{2}%
\theta\left( 1-\left\vert s\right\vert \right) .
\end{equation}
The next step is to make the integral dimensionless by introducing $\tau=st$
and dimensionless 2D vectors $\mathbf{x,y,z=p/}k_{F}\mathbf{,q/}k_{F}\mathbf{%
,Q}/k_{F}$. Using the expression for $H\left( q\right) $ in the classical
limit, $H\left( q\right) =\frac{T}{\rho\omega_{q}^{2}}$ ($\omega_{q}$ is
defined in Eq. (4m)), we obtain the decay function as given in Eq. (11m)%
\begin{equation}
F_{\phi}\left( t\right) =c_{\phi}^{2}tTf_{\phi}\left( \alpha
k_{F}^{2}t,Z^{-1/\eta}q_{c}/k_{F}\right) \frac{T}{\mu}.   \label{f3}
\end{equation}
Here,%
\begin{equation}
f_{\phi}\left( \mathcal{T},\xi\right) =4\pi^{2}\int\left( 2\pi\right)
^{2}\left( d\mathbf{z}\right) \left( d\mathbf{x}\right) \left( d\mathbf{y}%
\right) \delta\left( \mathbf{z}-\mathbf{x-y}\right) S\left( z\right) \frac{%
\left( \mathbf{x}\cdot\mathbf{y}\right) ^{2}}{x^{4}y^{4}}\int\Xi_{-}\left(
s\right) ds\frac{\cos\left( \mathcal{T}sx^{2}\Theta _{\xi}\left( x\right)
\right) \cos\left( \mathcal{T}sy^{2}\Theta_{\xi }\left( y\right) \right) }{%
\Theta_{\xi}^{2}\left( x\right) \Theta_{\xi }^{2}\left( y\right) } 
\label{f1a}
\end{equation}
with $\mathcal{T}=\alpha k_{F}^{2}t,$ $\xi=Z^{-1/\eta}q_{c}/k_{F}$, $%
\Theta_{\xi}\left( x\right) =\sqrt{1+\left( x/\xi\right) ^{-\eta}}$, and 
\begin{equation}
S_{\phi}\left( z\right) =\left( \frac{z}{1+z/\left( g_{e}N\right) }\right)
^{2}\sqrt{1-\left( z/2\right) ^{2}}\theta\left( 2-z\right) ,   \label{sx}
\end{equation}
which is the product of the screening and chiral factors. Here $N=4$ is the
spin-valley degeneracy in graphene, and $g_{e}$ describes the renormalized
Coulomb interaction.

To proceed, it is convenient to make some transformations in Eq. (\ref{f1a}%
). First, we integrate out $s$ exactly, using the relation%
\begin{equation}
\int\Xi_{\pm}\left( s\right) ds\cos\left( sa\right) \cos\left( sb\right)
=\Xi_{\pm}\left( a,b\right) ,
\end{equation}
with\qquad\ 
\begin{equation}
\Xi_{\pm}\left[ a,b\right] =1\pm\frac{a\sin a\cos b-b\sin b\cos a}{%
a^{2}-b^{2}}.
\end{equation}
This gives%
\begin{equation}
f_{\phi}\left( \mathcal{T},\xi\right) =\left( 2\pi\right) ^{4}\int\left( d%
\mathbf{z}\right) \left( d\mathbf{x}\right) \left( d\mathbf{y}\right)
\delta\left( \mathbf{z}-\mathbf{x-y}\right) S_{\phi}\left( z\right) \frac{%
\left( \mathbf{x}\cdot\mathbf{y}\right) ^{2}}{x^{4}y^{4}}\frac{\Xi _{-}\left[
\mathcal{T}\Theta_{\xi}\left( x\right) x^{2}s,\mathcal{T}\Theta_{\xi}\left(
y\right) y^{2}s\right] }{\Theta_{\xi}^{2}\left( x\right)
\Theta_{\xi}^{2}\left( y\right) }.   \label{f1b}
\end{equation}
The final step is to resolve the delta-function in the above expression for $%
\mathbf{y}$ which yields 
\begin{equation}
y=\sqrt{z^{2}+x^{2}-2zx\cos\psi},   \label{fy}
\end{equation}
where $\psi$ is the angle between $\mathbf{z}$ and $\mathbf{x}$. As a
result, one finally obtains%
\begin{equation}
f_{\phi}\left( \mathcal{T},\xi\right) =\int_{0}^{2}dzS_{\phi}\left( z\right)
\int_{0}^{\infty}dx\int_{0}^{2\pi}\frac{d\psi}{2\pi}\frac{\left(
z\cos\psi-x\right) ^{2}}{xy^{4}}\frac{\Xi_{-}\left[ \mathcal{T}\Theta_{\xi
}\left( x\right) x^{2},\mathcal{T}\Theta_{\xi}\left( y\right) y^{2}\right] }{%
\Theta_{\xi}^{2}\left( x\right) \Theta_{\xi}^{2}\left( y\right) }, 
\label{f5}
\end{equation}

\subsection{Vector potential coupling}

We define 
\begin{equation}
F_{A}^{l}\left( t\right) \equiv c_{A}^{2}tTf_{A}^{l}\left( \alpha
k_{F}^{2}t,Z^{-1/\eta}q_{c}/k_{F}\right) \frac{T}{\mu},   \label{fa1}
\end{equation}
where $c_{A}$ is the dimensionless el-FP coupling constant for the vector
potential.

\subsubsection{Intervalley channels $\left( l=0,z\right) $}

Since $s_{l}=-1$, the decay function for the intervalley Cooperons is
similar with that for the scalar potential. Without providing further
details, we conclude that 
\begin{equation}
f_{A}^{l}\left( \mathcal{T},\xi\right) =\int_{0}^{2}dzS_{A}\left( z\right)
\int_{0}^{\infty}dx\int_{0}^{2\pi}\frac{d\psi}{2\pi}\frac{1}{xy^{2}}\frac {%
\Xi_{-}\left[ \mathcal{T}\Theta_{\xi}\left( x\right) x^{2},\mathcal{T}%
\Theta_{\xi}\left( y\right) y^{2}\right] }{\Theta_{\xi}^{2}\left( x\right)
\Theta_{\xi}^{2}\left( y\right) },   \label{fainter}
\end{equation}
where $y$ has been defined by Eq. (\ref{fy}), and 
\begin{equation}
S_{A}\left( z\right) =\frac{1}{2}\left( 1-\left( z/2\right) ^{2}\right)
^{-1/2}\theta\left( 2-z\right) .
\end{equation}
Note that $S_{A}\left( z\right) $ includes only the chiral factor since
screening does not affect the vector potential coupling constant.

\subsubsection{Intravalley channels $\left( l=x,y\right) $}

For the intravalley channels, $s_{l}=1$. Correspondingly $\Xi_{-}$ has to be
changed to $\Xi_{+}$:%
\begin{equation}
f_{A}^{l}\left( \mathcal{T},\xi\right) =\int_{0}^{2}dzS_{A}\left( z\right)
\int_{0}^{\infty}dx\int_{0}^{2\pi}\frac{d\psi}{2\pi}\frac{1}{xy^{2}}\frac {%
\Xi_{+}\left[ \mathcal{T}\Theta_{\xi}\left( x\right) x^{2},\mathcal{T}%
\Theta_{\xi}\left( y\right) y^{2}\right] }{\Theta_{\xi}^{2}\left( x\right)
\Theta_{\xi}^{2}\left( y\right) }.
\end{equation}

\section{Asymptotic properties of the decay function and the phase diagram}

The analytical expressions obtained in the previous sections allow us to
determine $\tau_{\phi/A}$ at arbitrary temperature $T$ and chemical
potential $\mu$.\ (In the end, the dephasing times $\tau_{\phi/A}$ are
defined as solution of the equations $F_{\phi/A}\left( \tau_{\phi/A}\right)
=1.$) We have developed numerical procedure, which exploits these equations
to calculate corresponding $\tau_{\phi}\left( T,\mu\right) $ dependencies.
Before presenting general results, let us concentrate on the properties of
the functions $f_{\phi/A}\left( \mathcal{T},\xi\right) ~$in the analytically
accessible regimes.

\subsection{Scalar potential coupling}

\subsubsection{Large $\protect\xi$}

For $\xi\gg1$, one may identify three asymptotic regions depending on $%
\mathcal{T}$. When $\mathcal{T}\ll\xi^{-2}$, the integral in $f_{\phi}\left( 
\mathcal{T},\xi\right) $ is dominated by the quadratic spectrum where $%
\Theta_{\xi}\left( x\right) =1$ (refer to Eq. (\ref{f5})), while at $%
\xi^{-2}\ll\mathcal{T}\ll\xi^{-\eta/2}$ the integral is dominated by the
part of spectrum where $\Theta_{\xi}\left( x\right) =\left( x/\xi\right)
^{-\eta/2}$. Thus, for not too large $\mathcal{T}$ $\left( \mathcal{T}\ll
\xi^{-\eta/2}\right) $, one may assume that the spectrum is homogeneous, and
put $\Theta_{\xi}\left( x\right) =\left( x/\xi\right) ^{b}$. Then the
function $f_{\phi}$ reads:%
\begin{equation}
f_{\phi}\left( \mathcal{T},\xi\right) =\frac{\xi^{4b}}{2\pi}%
\int_{0}^{2}dzS_{\phi}\left( z\right)
\int_{0}^{\infty}dx\int_{0}^{2\pi}d\psi \frac{\left( z\cos\psi-x\right) ^{2}%
}{x^{1+2b}y^{4+2b}}\Xi_{-}\left[ \mathcal{T}\left( x/\xi\right) ^{b}x^{2},%
\mathcal{T}\left( y/\xi\right) ^{b}y^{2}\right] .   \label{scaling}
\end{equation}
We obtain,%
\begin{equation}
f_{\phi}\left( \mathcal{T},\xi\right) =c\left( b\right) \xi^{\frac {6b}{2+b}}%
\mathcal{T}^{\frac{2+4b}{2+b}},   \label{sc}
\end{equation}
with $c\left( b\right) =\frac{M_{b}N_{\phi}}{2+b},$ where $%
N_{\phi}=\int_{0}^{2}dzS_{\phi}\left( z\right) =0.89$ (assuming $%
g_{e}\approx1$), and 
\begin{equation}
M_{b}=\int_{0}^{\infty}u^{-\frac{5b+4}{b+2}}\Xi_{-}\left( u,u\right) du=2^{%
\frac{3b}{b+2}}\Gamma\left( -\frac{5b+4}{b+2}\right) \sin\frac{\left(
5b+4\right) \pi}{\left( b+2\right) 2}.   \label{mb}
\end{equation}
More specifically, for $b=-\eta/2,$ $c_{I}^{\eta}\equiv c\left(
-\eta/2\right) =0.03$, when $\eta=0.8$. For $b=0$, $c_{II}\equiv c\left(
0\right) =\frac{\pi}{8}N_{\phi}$. The scaling expression in Eq. (\ref{sc})
can be achieved due to the fact that one can neglect all the $z$ dependences
except for $S_{\phi}\left( z\right) $ in Eq. (\ref{scaling}). Then, $y\simeq
x$, $\psi$ integral gives $2\pi$, and the remaining integral over $x$ leads
to Eq. (\ref{sc}). The obtained asymptotes will describe non-golden rule
behavior for the decay function, in the region I $\left( b=-\eta/2\right) $
and II $\left( b=0\right) $. Note that for $b=0,$ $f_{\phi}=c_{II}\mathcal{T}
$.

Next, for $\mathcal{T}\gg\mathcal{\xi}^{-\eta/2}$, the golden rule regime
holds and the vertex diagram is not important any more. In this case, one
can put $\Xi_{-}\rightarrow1$, and $f_{\phi}\left( \mathcal{T},\xi\right) $
becomes independent on $\mathcal{T}$:

\begin{align}
f_{\phi}\left( \mathcal{T},\xi\right) & =\int_{0}^{2}dzS_{\phi}\left(
z\right) \int_{0}^{\infty}dx\int_{0}^{2\pi}\frac{d\psi}{2\pi}\frac{\left(
z\cos\psi-x\right) ^{2}}{xy^{4}}\frac{1}{\Theta_{\xi}^{2}\left( x\right)
\Theta_{\xi}^{2}\left( y\right) }  \notag \\
& =c_{IV}^{\eta}\xi^{-2\eta},   \label{fIV}
\end{align}
where $c_{IV}^{\eta}=\eta U^{\eta}W^{\eta}\int_{0}^{2}dzS_{\phi}\left(
z\right) z^{2\eta-2}.$ Here, with $J_{n}\left( u\right) $ being the Bessel
function of the first kind, $U^{\eta}\equiv\int_{0}^{\infty}u^{-2\eta}J_{1}%
\left( u\right) du=\frac{4^{-\eta}\Gamma\left( 1-\eta\right) }{\Gamma\left(
1+\eta\right) },$ and $W^{\eta}\equiv\left( \int_{0}^{\infty
}u^{\eta-1}J_{0}\left( u\right) du\right) ^{2}+\left( \int_{0}^{\infty
}u^{\eta-1}J_{2}\left( u\right) du\right) ^{2}$ $=\left( \frac{2^{\eta
-1}\Gamma\left( \eta/2\right) }{\Gamma\left( 1-\eta/2\right) }\right)
^{2}+\left( \frac{2^{\eta-1}\Gamma\left( 1+\eta/2\right) }{\Gamma\left(
2-\eta/2\right) }\right) ^{2}$. For $\eta=0.8,$ this yields $U^{\eta}=1.63$
and $W^{\eta}=2.42.$ Eq. (\ref{fIV}) describes the asymptotic behavior of
the function $f_{\phi}\left( \mathcal{T},\xi\right) $ in region IV.

As a result, we see that the function $f_{\phi}\left( \mathcal{T},\xi\right) 
$ for large $\xi$ evolves with growing $\mathcal{T}$ as follows (regions of
applicability can be easily read off from these equations):

\begin{equation}
c_{II}\mathcal{T}\rightarrow c_{I}^{\eta }\xi ^{-\frac{3\eta }{2-\eta /2}}%
\mathcal{T}^{\frac{2-2\eta }{2-\eta /2}}\rightarrow c_{IV}^{\eta }\xi
^{-2\eta }.  \label{largexi}
\end{equation}%
As indicated in the main text, the matching of different asymptotes gives us
a separation between different regions of the dephasing rate on the $\left(
T-\mu \right) $ plane: The black line in the Fig. 3m corresponds to matching
of the first pair of asymptotes. The location of the region IV is related to
matching of the second pair.

\subsubsection{Small $\protect\xi$}

For small $\xi \ll 1$, there are two asymptotic regions depending on the
value of $\mathcal{T}$. For $\mathcal{T}\ll 1,$ the integral in Eq. (\ref%
{scaling}) is always dominated by the quadratic spectrum ($b=0$), and
function $f_{\phi }\left( \mathcal{T},\xi \right) $ acquires the asymptote
linear in $\mathcal{T}$ (compare with Eq. (\ref{sc}) at $b=0$):%
\begin{equation}
f_{\phi }\left( \mathcal{T},\xi \right) =c_{II}\mathcal{T}.  \label{cIIT}
\end{equation}%
At large $\mathcal{T}\gg 1$, the golden rule is applicable. In this region, 
\begin{equation}
f_{\phi }\left( \mathcal{T},\xi \right) =c_{III}\log 1/\xi ,
\label{xismallsc0}
\end{equation}%
where $c_{III}=\int_{0}^{2}S_{\phi }\left( z\right) /z^{2}dz$. Note, that $%
c_{I}^{\eta }$, $c_{II}$, $c_{III}$ and $c_{IV}^{\eta }$ implicitly depend
on $g_{e}$ via $S_{\phi }\left( z\right) $, see Eq. (\ref{sx}). Thus, in the
case of small $\xi $, $f_{\phi }\left( \mathcal{T},\xi \right) $ evolves
with growing $\mathcal{T}$ as follows%
\begin{equation}
c_{II}\mathcal{T}\rightarrow c_{III}\log 1/\xi .  \label{xismallsc}
\end{equation}%
The blue line in the Fig. 3m corresponds to matching of the above
asymptotes. 

Equation (\ref{f5}) allows for straightforward numerical evaluation. Here as
an illustration, we present the numerical calculation of the function $%
f_{\phi}\left( \mathcal{T}\right) $. From the plot we see an excellent
agreement with theoretical calculation for the region $\mathcal{T}\ll1$ and $%
\mathcal{T}\gg1$, even for an intermediate value of $\xi=0.52$. 
\begin{figure}[tbh]
\centering
\includegraphics[width=.4\textwidth]{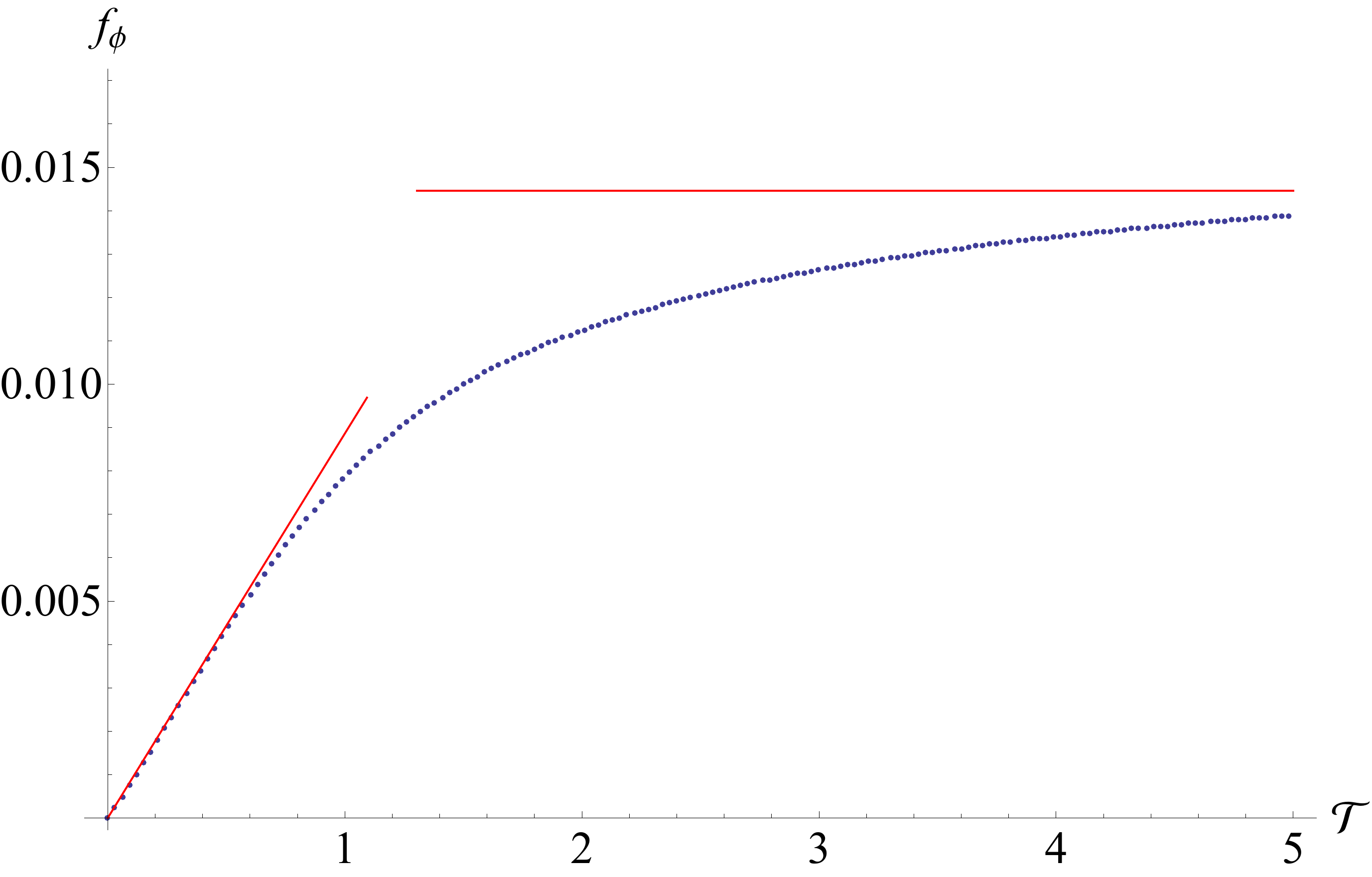}
\caption{Plot of the function $f_{\protect\phi}\left( \mathcal{T}\right) $.
The parameters are taken as follows: $\protect\eta=0.8,$ $\protect\xi=0.52$,
and $Z=3$. The blue dots are numerical calculation and the red lines are the
asymptotes for different region mentioned above.}
\label{asym}
\end{figure}

\subsubsection{Phase diagram of the dephasing rate}

\begin{figure}[tbh]
\centering
\includegraphics[width=.4\textwidth]{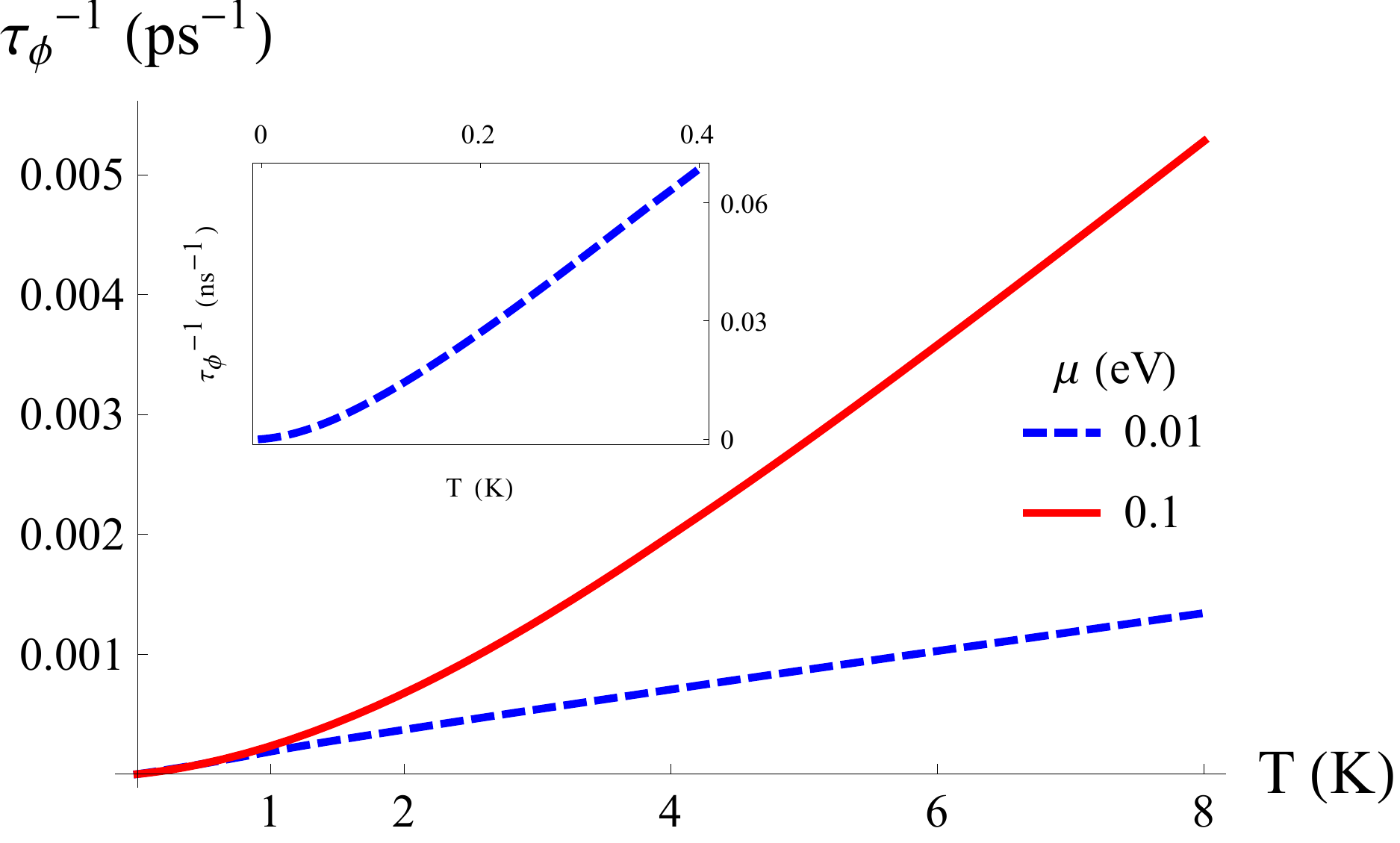}
\caption{Depahsing rate as a function of temperature at two different chemical potentials.
The transition from a quadratic to linear dependence takes place at $\sim$4K for the red solid line. 
For lower chemical potential (represented by the blue dashed line), the transition happens at a lower temperature 
as shown in the inset.}
\label{temp}
\end{figure}

In Fig. 3m, the blue and black lines have been found by matching the
asymptotic behavior of the dephasing rates deep in regions I, II and III.
The blue line is found as a consequence of matching Eq. (\ref{cIIT}) with
Eq. (\ref{xismallsc0}). It is defined by the criterion $\omega _{k_{F}}\tau
_{\phi }\approx 3,$ where $\tau _{\phi }$ is calculated numerically from Eq.
(11m) with the phonon frequency $\omega _{k_{F}}$ taken from Eq. (4m). The
black line is found by matching the first pair in Eq. (\ref{largexi}), and
it can be obtained according to the equation $\omega _{q_{c}}\tau _{\phi
}\approx 7$. As mentioned in the main text, the obtained rate has a transition from $\tau_{\phi}^{-1} \sim T^2$
to $\tau_{\phi}^{-1} \sim T$ when temperature is increased. This transition is illustrated on Fig. \ref{temp}.

Note that besides the regions discussed in the main text, on the $\left(
T-\mu\right) $ plane there exists another asymptotic region (IV), not
mentioned in Eqs. (12m). It lies at very small densities below the blue line
and above the line $\xi=1$. In this region, the dephasing rate is still
described by the GR, but contrary to region III, renormalization of the FP
spectrum at low momenta is essential. In this situation, the dephasing rate
equals $\tau_{\phi}^{-1}\sim\gamma T\left(\mu/\mu_{0}\right) ^{2\eta
-1}\left( T/T_{0}\right) ^{1-\eta}$. However, the relevant densities are so
small, that this region does not fit into the scale of Fig. 3m.

It is instructive to compare
the situation with flexural phonons to that for ordinary phonons in semiconductors, where the
dephasing process can be viewed as coming from the energy diffusion via
low-energy (i.e., quasi-elastic) collisions \cite{Altshuler81}. Under these
conditions, the dephasing rate $\tau_{\phi}^{-1}\sim\sqrt{\left(  \tau_{\phi
}/\tau_{out}\right)  \delta\epsilon^{2}}$, where $\delta\epsilon$ denotes the
characteristic energy transfer during a single scattering event. Thus, the
accumulation of phase in the course of the energy diffusion yields $\tau
_{\phi}^{-1}\sim\left(  \delta\epsilon^{2}/\tau_{out}\right)  ^{1/3}$.
The specific point of our problem, compared to dephasing via usual phonons, is
that an electron is coupled to two flexural phonons, and dephasing cannot be
described by the energy diffusion process. Because of this \textit{two}-phonon
interaction, the support of the correlation function of the fluctuations is
not characterized by any typical frequency $\Omega$, and frequency transfer
occurs
in such a way that
typical energy transfer is of the order of $\tau_{\phi}^{-1}(T)$. In
particular, this allows for an energy transfer exceeding $T_{BG}$, although
the momentum transfer is limited by $2k_{F}$. As a result, the decay function
$F_{\phi}(t)$ at short times (i.e., in the non-GR regime) is proportional to
$t^{2}$, rather than $t^{3}$ as for the case of the energy diffusion. Note
that the energy diffusion corresponds to the expansion of the factor
$\mathcal{C}^{\phi}(t)\sim(\Omega t)^{2}$ which leads to the $t^{3}%
$-dependence of $F_{\phi}(t)$. As we have already mentioned, in the case of
FPs one cannot expand $\mathcal{C}^{\phi}(t)$.

\subsection{Vector potential coupling}

\subsubsection{Intervalley channel}

The similarity (compare Eq. (\ref{f5}) with Eq. (\ref{fainter})) with the
scalar potential case makes the calculation of asymptotes for the function $%
f_{A}^{l}\left( \mathcal{T},\xi\right) $ straightforward. The function $%
f_{A}^{l}\left( \mathcal{T},\xi\right) $ evolves with the growth of $%
\mathcal{T}$ in the following fashion:

for $\xi\gg1,$%
\begin{equation}
d_{II}\mathcal{T}\rightarrow d_{I}^{\eta}\xi^{-\frac{3\eta}{2-\eta/2}}%
\mathcal{T}^{\frac{2-2\eta}{2-\eta/2}}\rightarrow d_{IV}^{\eta}\xi^{-2\eta};
\end{equation}

while for $\xi\ll1,$%
\begin{equation}
d_{II}\mathcal{T}\rightarrow d_{III}\xi^{-1}.   \label{xismall}
\end{equation}
Note that due to the absence of screening, dependence on $\xi$ is stronger
than in the case of the scalar potential, compare Eq. (\ref{xismallsc}) with
Eq. (\ref{xismall}).

Let us present some technical details. Here $d_{I}^{\eta}\equiv d\left(
-\eta/2\right) ,$ where $d\left( b\right) =\frac{\pi}{4}\frac{M_{b}}{1+b/2}$
and $M_{b}$ is defined in Eq. (\ref{mb}); specifically, for $b=0,$ $%
d_{II}\equiv d\left( 0\right) =1/64$. In the GR region, the function $%
f_{A}^{l}\left( \mathcal{T},\xi\right) $ is independent on $\mathcal{T}$.
Its asymptotic behavior is given as follows:%
\begin{align}
f_{A}\left( \xi\right) & =\int_{0}^{2}dzS_{A}\left( z\right)
\int_{0}^{\infty}dx\int_{0}^{2\pi}\frac{d\psi}{2\pi}\frac{1}{xy^{2}}\frac {1%
}{\Theta_{\xi}^{2}\left( x\right) \Theta_{\xi}^{2}\left( y\right) }  \notag
\\
& \equiv\left\{ 
\begin{array}{ccc}
d_{III}\xi^{-1} & \text{for} & \xi\ll1 \\ 
d_{IV}^{\eta}\xi^{-2\eta} & \text{for} & \xi\gg1%
\end{array}
\right. .   \label{d2}
\end{align}
Here, for $\eta>1/2$ (which is important for the convergence of the
following integral),%
\begin{equation}
d_{III}=\frac{1}{2}\int_{0}^{\infty}dz\int_{0}^{\infty}dx\int_{0}^{2\pi}%
\frac{d\psi}{2\pi}\frac{1}{xy^{2}}\frac{1}{\Theta_{\xi=1}^{2}\left( x\right)
\Theta_{\xi=1}^{2}\left( y\right) },
\end{equation}
and%
\begin{equation}
d_{IV}^{\eta}=2\eta U^{\eta}V^{\eta}G_{A},
\end{equation}
where $U^{\eta}$ is the same as in the case of the scalar potential, $%
V^{\eta }\equiv\left( \int_{0}^{\infty}u^{\eta-1}J_{0}\left( u\right)
du\right) ^{2}=\left( \frac{2^{\eta-1}\Gamma\left( \eta/2\right) }{%
\Gamma\left( 1-\eta/2\right) }\right) ^{2}$ and $G_{A}=\int_{0}^{2}dzS_{A}%
\left( z\right) z^{2\eta-2}=\frac{4^{\eta}\sqrt{\pi}}{8}\frac{\Gamma\left(
\eta-1/2\right) }{\Gamma\left( \eta\right) }$. For $\eta=0.8,$ this yields $%
V^{\eta}=1.68$ and $G_{A}=1.73$.

\subsubsection{Intravalley channel}

For the intravalley Cooperon, the function $f_{A}^{l}\left( \mathcal{T}%
,\xi\right) $ involves $\Xi_{+}$ which does not vanish in the limit of $%
\mathcal{T}\rightarrow0$. Therefore, it evolves with the growth of $\mathcal{%
T}$ (i.e., from $\mathcal{T}\ll1$ to $\mathcal{T}\gg1$) as

\begin{equation}
2f_{A}\left( \xi \right) \rightarrow f_{A}\left( \xi \right) ,
\end{equation}%
where $f_{A}\left( \xi \right) $ is defined in Eq. (\ref{d2}). It is
noteworthy that the static deformations produce dephasing here. This happens
because the electrons on the interfering trajectories are coupled to the
vector potential of the same sign. The situation is similar to that for a
particle in the random magnetic field \cite{Aronov94a,Aronov94b}.
Unfortunately, the observation of this effect is obscured by the gaps
inevitably produced in these channels by disorder scattering. The rate
obtained from $F_{A}^{l}\left( t\right) =1$ changes from $2\tau _{\ast
,A}^{-1}\ $to $\tau _{\ast ,A}^{-1}$ with increasing temperature, where 
\begin{equation}
\tau _{\ast ,A}^{-1}=\gamma T\left\{ 
\begin{array}{cc}
2.6\left( \mu /\mu _{0}^{\prime }\right) ^{2\eta -1}\left( T/T_{0}^{\prime
}\right) ^{1-\eta } & \text{for }\xi \gg 1 \\ 
1.9\sqrt{T/T_{0}^{\prime }} & \text{for }\xi \ll 1%
\end{array}%
\right. .
\end{equation}%
Here $\gamma \sim 0.02$ is the adiabatic parameter defined in the main text, 
$\mu _{0}^{\prime }\sim \gamma \Delta _{c}$ and $T_{0}^{\prime }\sim \gamma
^{2}\Delta _{c}$ where $\Delta _{c}$ describes the energy scale of the
anharmonicity (see the main text). Note that for $\xi \ll 1$, the
characteristic momentum transfer is $\sim $ $q_{c}\left( \equiv \frac{\sqrt{%
T\Delta _{c}}}{v_{F}}\right) $ instead of $k_{F}$. As a result, the rate is
independent of the chemical potential. Note also that for the intravalley
channels the diffusive limit $\left( q_{c}l\ll 1\right) $ for the
calculation of dephasing due to the el-FP interaction can be achieved at
higher temperature as compared to the intervalley channels. For example, for
dimensionless sheet conductance $g_{_{\square }}\sim 10$ and chemical
potential $\mu \sim 0.1$eV, the diffusive limit occurs already below $T\sim
0.5$K.

\bibliography{Flexural}